\documentclass[conference,a4paper]{IEEEtran}
\IEEEoverridecommandlockouts
\usepackage{amsmath,amssymb,amsfonts}
\usepackage{algorithmic}
\usepackage{graphicx}
\usepackage{textcomp}
\usepackage{xcolor}
\usepackage[colorinlistoftodos,prependcaption,textsize=tiny]{todonotes}

\usepackage{cite}
\usepackage{url}
\usepackage{hyperref}
\usepackage{comment}
\usepackage{theorem}
\usepackage{calligra}
\usepackage{array}

\newcommand{\hash}{\mathcal H}
\newcommand{\merkle}{\mathcal{M}}
\newcommand{\mroot}{\merkle.\textsf{Root}}

\usepackage{tikz}
\usetikzlibrary{arrows,automata, positioning}
\usetikzlibrary{calc, chains,
                fit,
                positioning,
                shapes}
\usetikzlibrary{patterns}       

\usepackage{pgfplots}

\usetikzlibrary{fadings,decorations.pathmorphing}

\usetikzlibrary{automata,arrows,positioning,calc}
\definecolor{myblue}{RGB}{0,0,205}
\definecolor{mygreen}{RGB}{80,160,80}
\definecolor{myred}{RGB}{178,34,34}
\definecolor{mygray}{RGB}{211,211,211}
\definecolor{myyellow}{rgb}{1.0, 0.75, 0.0}
\definecolor{mygreen2}{rgb}{0.4, 0.69, 0.2}
\definecolor{myred2}{rgb}{0.89, 0.15, 0.21}
\definecolor{myblue2}{rgb}{0.0, 0.47, 0.75}

\usepackage[boxed, linesnumbered, noend]{algorithm2e}

\SetCommentSty{mycommfont}

{\theorembodyfont{\rmfamily}
   }
{\theorembodyfont{\rmfamily}
   \newtheorem{Prop}{{\textbf Proposition}}}
{\theorembodyfont{\rmfamily}
   }
{\theorembodyfont{\rmfamily}
   }
{\theorembodyfont{\upshape}
   }
{\theorembodyfont{\upshape}
   }
{\theorembodyfont{\rmfamily}
   }

\usetikzlibrary{patterns}
\usepackage{scalerel}

\newcommand{\ie}{\textit{i.e.}}
\pgfplotsset{compat=1.17}
\begin{document}

\title{Optimization of a Reed-Solomon code-based protocol against blockchain data availability attacks
\\
}

\author{Paolo Santini, Giulia Rafaiani, Massimo Battaglioni, Franco Chiaraluce, and Marco Baldi\\

\IEEEauthorblockN{
\textit{Dipartimento di Ingegneria dell'Informazione}
\\{Università Politecnica delle Marche, Ancona (60131), Italy }
\\email:\{p.santini, g.rafaiani, m.battaglioni, f.chiaraluce, m.baldi\}@univpm.it}}

\maketitle

\begin{abstract}
ASBK (named after the authors' initials) is a recent blockchain protocol tackling data availability attacks against light nodes, employing two-dimensional Reed-Solomon codes to encode the list of transactions and a random sampling phase where adversaries are forced to reveal information.
In its original formulation, only codes with rate $1/4$ are considered, and a theoretical analysis requiring computationally demanding formulas is provided.
This makes ASBK difficult to optimize in situations of practical interest.
In this paper, we introduce a much simpler model for such a protocol, which additionally supports the use of codes with arbitrary rate. 
This makes blockchains implementing ASBK much easier to design and optimize.
Furthermore, disposing of a clearer view of the protocol, some general features and considerations can be derived (e.g., nodes behaviour in largely participated networks).
As a concrete application of our analysis, we consider relevant blockchain parameters and find network settings that minimize the amount of data downloaded by light nodes.
Our results show that the protocol benefits from the use of codes defined over large finite fields, with code rates that may be even significantly different from the originally proposed ones.
\end{abstract}

\begin{IEEEkeywords}
Blockchain, Data Availability Attack, Reed-Solomon Codes.
\end{IEEEkeywords}

\section{Introduction}
\label{introduction}
Given the recent blooming and spreading of blockchain applications, the scalability of existing networks based on this technology represents a serious issue.
Indeed, as the number of users grows,
an increasingly larger number of transactions must be handled.
However, the majority of existing blockchains impose some limit on the maximum dimension of each block, which translates into an upper bound on the number of transactions that can be validated per time unit.

According to the Simplified Payment Verification \cite{Nakamoto} paradigm, a light node only verifies the inclusion of  specific transactions and, hence, does not need to download the entire blockchain.
In such a scenario, \textit{full nodes}, that is, nodes with  more computational resources and full privileges, are the sole responsible for proposing and validating new blocks.
However, this distinction affects the decentralization and security of the network. In fact, as the number of validators decreases, the probability that the majority of them collude to create invalid transactions gets higher.  
This makes the straightforward solution of increasing the block size limits hardly recommendable.
In fact, this would cause an increase in the resources needed to store a copy of the ledger and, therefore, to run a full node capable of validating the blockchain. Consequently, users would more likely run \textit{light nodes}, which are not able to verify the transactions correctness.
Therefore, in a network with a majority of dishonest consensus-participating nodes, light nodes may not be able to detect invalid transactions included in a block by dishonest full nodes.
A solution to this problem consists in requiring honest full nodes to produce evidences of the invalidity of a transaction and to broadcast it to every connected light node, so that the latter becomes aware of the fraud and eventually rejects invalid blocks.

However, networks with a majority of dishonest full nodes are vulnerable to \textit{data availability attacks}. These attacks consist in a malicious node including invalid transactions in a block; the block header is then distributed to the network, but the malicious node withholds the part of the block containing invalid transactions. This way, a honest full node cannot validate the block and is also unable to demonstrate that a fraud attempt is occurring; in fact, in order to generate a fraud proof for an invalid block, it is necessary that all the transactions included in the block are available.

Hence, light nodes are interested to know whether all the data in a block are available to the network or not.
In order to contribute to countering these attacks, light nodes could randomly ask some pieces of a block and discard the entire block if they do not receive any answer. Instead, if their request is answered, they can forward the received data to the neighbouring full nodes and wait for a valid fraud proof. This way, malicious nodes may be required to release some part of the hidden information. Still, as the block size increases, malicious nodes could hide very small parts of the block, thus reducing the probability for light nodes to successfully sample withheld data. 
Some protocols based on linear codes have been proposed to address this issue \cite{asbk21, yu2020coded, Dolececk2020}.
In this paper, we focus on the protocol in \cite{asbk21}, called ASBK according to the authors' initials.
According to ASBK, the list of transactions of a block is encoded through two-dimensional Reed-Solomon (RS) codes, so that honest full nodes can recover all the data even when a relatively small portion of the block is missing. 
Hence, for preventing honest full nodes from recovering missing parts through decoding, malicious nodes should increase the portion of hidden data, thus increasing the probability for light nodes to sample it.

\paragraph*{Our contribution}{
We derive a simplified model for the coded blockchain protocol in \cite{asbk21} that allows us to deepen its analysis, with the aim of optimizing the involved parameters for minimizing the amount of data that light nodes have to download as well as the number of samples a light node should ask for.
We demonstrate that significant improvements can be achieved through our approach over the original one in \cite{asbk21}.

The paper is organized as follows. In Section \ref{background} we introduce the notation we use and provide the necessary background. In Section \ref{buterin} we describe and analyze the ASBK protocol. In Section \ref{implementation} we discuss our improved theoretical model, which is validated in Section \ref{results}. Section \ref{conclusions} concludes the paper.
} 

\section{Background}
\label{background}
In this section we recall basic concepts that are fundamental to our analysis.





\subsection{Coupon Collector's Problem}\label{sec:coupon}

The coupon collector's problem is a well-known classical problem in probability theory, and we refer to its formulation presented in \cite{wolfgang1990}.
Let us consider the set of coupon indexes $V = \{1,2,\ldots,v\}$ and a group of $y$ persons; each person picks a subset of $V$ containing $z\leq v$ distinct indexes.
Let $W\subseteq V$ of cardinality $w$ and $t = \left|W\cap\left(\bigcup_{i = 1}^y J_i\right)\right|$, where $J_i$ is the set of indexes selected by the $i$-th person.
When each $J_i$ is picked independently and uniformly at random, then $t$ is a random variable with mean value given by \cite{wolfgang1990}: 
\begin{equation}
\label{eq:mean_t}
\langle t \rangle = v\left(1-\left(1-\frac{z}{v}\right)^y\right).   
\end{equation}
The Cumulative Distribution Function (CDF) of $t$ results in
\begin{equation}
\label{eq:function_g}
\mathrm{Pr}[t < x] = \sum_{i = 0}^{x-1}(-1)^{x-i+1}\binom{w}{i}\binom{w-i-1}{w-x}\left(\frac{\binom{v-w+i}{z}}{\binom{v}{z}}\right)^y.
\end{equation}

\subsection{Reed-Solomon Codes}

Let $\mathbb F_q$ denote the finite field with $q$ elements. A linear code $\mathcal C$ defined over $\mathbb F_q$ with \textit{length} $n$ and \textit{dimension} $k$ is a $k$-dimensional subspace of $\mathbb F_q^n$. 
The \emph{code rate} is $R = k/n$.
Any linear code can be represented through a \emph{generator matrix} $\6G\in\mathbb F_q^{k\times n}$, such that each information vector $\6u\in\mathbb F_q^k$ is mapped into a codeword $\6c = \6u\6G$.
A special class of linear codes over $\mathbb F_q$ is that of RS codes.
An RS code $\mathcal C\subseteq \mathbb F_q^n$ with length $n\leq q$ and dimension $1\leq k < n$ is defined\footnote{
According to some authors (see, for instance, \cite[Chapter 10]{macwilliams}), RS codes are only defined with maximum length, i.e., $n = q-1$.
According to such a narrower definition, the codes described by \eqref{eq:rs_code} correspond to shortened RS codes, or to a special case of Generalized RS codes.
However, this little inconsistency in the nomenclature does not affect the properties of the family of codes we consider.
} as
\begin{equation}
\label{eq:rs_code}
\mathcal C = \left\{\big(g(a_1),\ldots,g(a_n)\big)\left|\hspace{1mm}g\in\mathbb F_q^k[x]\right.\right\},
\end{equation}
where $\{a_1,\ldots,a_n\}$ are distinct elements of $\mathbb F_q$ and $\mathbb F_q^k[x]$ is the set of polynomials with coefficients over $\mathbb F_q$ and maximum degree $k-1$.
In our model, malicious parties intentionally introduce \textit{erasures}: an erased symbol assumes an unknown value, and cannot be singularly recovered by the receiver. 
It can be proven that any code defined by \eqref{eq:rs_code} can fill up to $n-k$ erasures, \ie, any codeword can be recovered from any set of $k$ of its non-erased symbols using a conventional decoding algorithm \cite{Lin2004}. 
Normalizing with respect to the code length, we denote with $\gamma = \frac{k}{n} = R$ the minimum fraction of non-erased symbols allowing for full codeword recovery.


\subsection{Blockchain Technology}

The core of the blockchain technology is a distributed ledger having the form of a chain of blocks, where the link between adjacent blocks is obtained through cryptographic functions.
Each block contains an ordered list of data units called \textit{transactions}, except for the first one, which is fixed according to the particular blockchain and initiates the ledger. 
The blockchain is driven by a \textit{consensus protocol}, that sets all the rules to verify a block and append it to the chain. 
A validator chooses a set of valid transactions to be included in a block and proposes it to the network according to the consensus protocol. 
The transactions included in the block are used as leaves to build a Merkle tree. The root of the Merkle tree (\ie, the Merkle root) is stored, together with some additional information, in the block header.

We distinguish between \textit{full nodes} and \textit{light nodes}.
Full nodes participate in the blockchain with all the rights and duties. 
Namely, they are able to download and store the entire ledger and can actively participate in the consensus mechanism (\ie, proposing and validating blocks). 
On the contrary, light nodes do not have enough resources to store a copy of the entire ledger and/or to participate in consensus. 
For this reason, light nodes have a somehow limited capacity to interact with the network.
In fact, they do not participate in consensus (\ie, they cannot propose and validate blocks), and store only the block headers (instead of the full blockchain). 
Light nodes cannot autonomously verify the validity of transactions, and are only interested in verifying its inclusion in a valid block; they accomplish this task by means of Merkle proofs.
\subsection{Data Availability Attacks}

Let us consider the situation in which a block is not available to all the full nodes participating in the network.
This may be due to one of two possible reasons: either the block is valid but, due to fluctuations in the network synchronization, it has not still been received by all the network nodes, or the block is not valid, and the invalid portions have been withheld by malicious nodes. Light nodes do not have a reliable way to distinguish between these two cases. 
So, a malicious full node can take advantage of this situation and try to deceive the network, which happens when: i) the contents of a block are not available to the honest full nodes, and ii) at least one light node accepts the block.
Also notice that, in such a case, the malicious node can put invalid transactions in the unrevealed part; so, it can ultimately make a light node accept an invalid transaction.
In principle, honest full nodes could raise an alarm any time some part of a block is missing, but this could occur every time network synchronization slows down, with the consequence of flooding the network with many false alarms.
Moreover, malicious full nodes could raise fake alarms, preventing light nodes from accepting valid blocks.

In order to tackle these attacks, a better solution consists in relying on \textit{fraud proofs}, \ie, objects produced by full nodes to prove that a certain header is associated to an invalid block.
Upon receiving a fraud proof, light nodes become aware of the fraud and reject the block.
We describe next the ASBK protocol, which has been the first one to use fraud proofs.

\section{The ASBK protocol}
\label{buterin}

Let us consider the same threat model and network topology as in \cite{asbk21}.
In particular, we assume that the majority of full nodes is dishonest and produces invalid blocks.
The network is supposed to be reliable and partially asynchronous (\ie, the maximum delay is finite), with peer-to-peer authenticated communications.
Full nodes can communicate among themselves and with light nodes; additionally, we assume that malicious nodes can collude.
Light nodes cannot communicate among themselves, but can query full nodes with completely anonymized requests. 
Such an assumption, which in \cite{asbk21} is referred to as \textit{enhanced model}, is crucial to the scheme functioning, and involves that a malicious node  receives all the requests together and in mixed order, without any information about the sender. Concerning the topology, we assume that each light node is connected to at least one honest full node, while every full node is connected to $m$ light nodes.

\subsection{Protocol Description}



We generalize the setting in \cite{asbk21} and  consider RS codes having whichever value of code rate.
The ASBK protocol works by encoding the list of transactions in each block as a codeword $\6c$ of a 2D-RS code. 
In a nutshell, 2D encoding is performed through a product code, employing as component codes two identical RS codes with length $n'$, dimension $k'$ and rate $R'$, defined over a finite field with $q \geq n'$ elements. Hence, the resulting product code has length $n = n'^{2}$, dimension $k = k'^2$, and, consequently, rate $R = R'^2$.
It can be easily proven that such a construction yields a code that can recover up to $(n'-k'+1)^2-1$ erasures\footnote{We remark that the analysis in \cite{asbk21} is restricted to the case of $R'=\frac{1}{2}$ and therefore, as shown in \cite[Theorem 1]{asbk21}, the code is able to fill up to $(k'+1)^2-1$ erasures.}, and hence 
\begin{align}
\gamma &\nonumber = \frac{n'^2 - (n'-k'+1)^2+1}{n'^2}
= 1-\left(1-R'+1/n'\right)^2+\frac{1}{n'^2}\\
& \approx 1-\left(1-R'\right)^2 = R'(2-R').
\label{eq:gamma}
\end{align}

After receiving the header of a new block, light nodes start querying the connected full nodes, asking for random symbols of $\6c$ together with the Merkle proof.
Each received  symbol is then gossiped to the connected full nodes.
This way, honest full nodes will receive further entries of $\6c$ and, upon reception of enough symbols, they become able to retrieve the whole block through RS decoding: in case the retrieved block includes invalid transactions, they will produce the fraud proof and deliver it to light nodes, that will consequently reject the block.
Thus, in order to prevent honest full nodes from retrieving the whole block through RS decoding, malicious nodes neglect, \ie, do not reply to, some of the light nodes queries. This is the only possibility for malicious nodes to deceive the network.
Indeed, the light nodes that do not receive an answer will put the block in a pending state, but all the other light nodes (\ie, the ones for which all queries are replied) will accept the block, causing a fork in the blockchain.

In the enhanced network model, all queries are anonymous: malicious nodes choose the requests to be neglected only on the basis of the asked symbols, and not of the sender.
This means that there is still some probability that malicious nodes are able to deceive the network.
We call such a probability \textit{adversarial error probability}, and describe in the following section how it can be computed.



\subsection{Protocol Analysis}

Let us recall the analysis in \cite{asbk21} to estimate the adversarial success probability.
We remind that adversaries succeed whenever full nodes are unable to recover an invalid block through decoding and, at the same time, there is at least one light node that accepts the  block, since all its queries have been replied. 

Depending on the symbols asked by light clients, malicious nodes will behave differently, with the purpose of maximizing the probability to deceive the network. Let $J_i$ denote the set of indexes of symbols asked by the $i$-th light node, and let $J = \bigcup_{i = 1}^m J_i$.
Also, let $E$ be the set of indexes of the symbols which have been initially hidden by malicious nodes; we denote $|E| = \beta n$, where $\beta \in [\![ 0 ; 1 ]\!]$, \ie, the set of rational numbers between $0$ and $1$.
The fraction of symbols available to full nodes at the end of the sampling process, if the malicious node replied to all queries, is given by $\varphi n$, with
\begin{equation}
\label{eq:decode_coordinates}
\varphi = 1-\beta  + \frac{\left|J\cap E\right|}{n},
\end{equation}
which is easily obtained by summing the fraction $1-\beta$ of symbols already known to full nodes to the fraction of symbols obtained by light nodes through sampling.
The value of $\varphi$, which can be computed by the adversaries on the run, determines their behavior, namely:
\begin{itemize}
    \item[A) ] if $\varphi < \gamma$, full nodes will not be able to decode: the malicious node will reply to all queries, and all light nodes will accept the block;
    \item[B) ] if $\varphi \geq \gamma$, the malicious node must avoid replying to some queries. 
    Let $d$ be the minimum number of indexes from $E\cap J$ which, if not revealed, would make the block undecodable.
    It can be easily seen that it must be $d = \delta n + 1$, where $\delta = \varphi - \gamma$.  Note that, by adapting their behavior to the light nodes requests, malicious nodes maximize the probability to have at least one light node for which all queries are correctly replied.
\end{itemize}
The sampling process can be seen as an instance of the coupon's collector problem in which the variables $(v,y,z,w)$, defined in Section \ref{sec:coupon}, take the values $(n,m,s,\beta n)$.
Indeed, there is a group of $m$ light nodes, each asking for $s$ distinct symbols with indexes in $\{1,\ldots,n\}$; in our case, the set $W$ corresponds to that of hidden symbols, thus $w = \beta n$.
Hence, in order to estimate the probability associated to condition A), it is enough to plug $x = (\gamma+\beta-1) n - 1$ into \eqref{eq:function_g}.\footnote{The authors in \cite{asbk21} provide a formula (see Theorem 4) which is essentially the same, apart from some rewriting. 
We have chosen a different formulation only to simplify the treatment in the subsequent computations.}
In the following, we will call $\mathrm{Pr}[\textsf{NoDec}]$ the resulting probability.

Moreover, in order to assess the adversarial success probability, one should also take into account condition B), which happens with probability $1 - \mathrm{Pr}[\textsf{NoDec}]$.
Then, we have to consider the probability that there is at least one light node that receives all the asked symbols, which we are going to denote as $\mathrm{Pr}[\textsf{Deny}]$. Putting all of this together, we finally obtain that the adversarial success probability is
\begin{equation}
\label{eq:eps}
\epsilon = \mathrm{Pr}[\textsf{NoDec}] + \left(1-\mathrm{Pr}[\textsf{NoDec}]\right)\mathrm{Pr}[\textsf{Deny}].
\end{equation}


\section{A General Analysis of the ASBK scheme}
\label{implementation}
In this section, we describe how the analysis proposed in \cite{asbk21} can be simplified and generalized.
We still consider 2D-RS codes, but with variable rate $R\in [\![ 0 ; 1 ]\!]$ as a parameter to optimize.
In addition, we get rid of  hard-to-implement formulas, in favour of a simpler and more intuitive analysis.

We first notice that, in order to implement \eqref{eq:function_g} in our case,  a significant amount of computational resources is needed.
Indeed, a rigorous computation of  \eqref{eq:function_g} requires a number of operations which is $O\left(n^2 + n^2\log_2(m)\right)$, having assumed that computing a binomial $\binom{a}{b}$ costs $O(b)$ operations, and that computing $a^b$ costs $O\big(\log_2(b)\big)$ operations. 
Considering that we expect to have $n \sim 10^4 \div 10^5$ and that we need to numerically search for optimal parameters, we observe that performing these computations may be not easy.
Note that there exist ways to speed-up the process: for instance, one can approximate binomials and can also rely on precomputation to reduce the number of operations to execute on the run.
Still, the computational burden remains significant, which motivates the need for a simpler analytical model like the one described in the next section.

\subsection{Simple Theoretical Model}

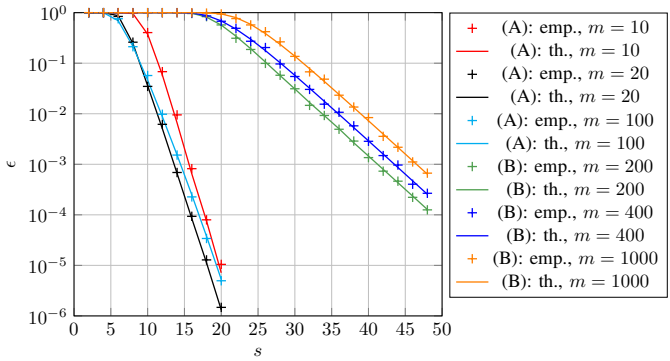
\begin{figure}
    \centering
    \resizebox{\columnwidth}{!}{%
    \begin{tikzpicture}
  \begin{axis}[
xtick={0,5,...,50},            ytick={1,0.1,0.01,0.001,0.0001,0.00001,0.000001,0.0000001},
xmin = 0,
xmax= 50,
grid = major,
ymin=10^-6,
ymax=1,
ymode = log,
legend style={at={(1.02,1)},anchor=north west},
mark size=2pt,
xlabel={$s$},
ylabel={$\epsilon$},
xticklabel style={
        /pgf/number format/fixed,
        /pgf/number format/precision=5
}]

\addplot[only marks, mark=+,red, mark size=2.5pt,line width=0.7pt]coordinates{
(2, 1.00000000000000)
(4, 1.00000000000000)
(6, 1.00000000000000)
(8, 0.980392156862745)
(10, 0.403225806451613)
(12, 0.0680272108843537)
(14, 0.00948766603415560)
(16, 0.000818732601932209)
(18, 0.0000796529679492388)
(20, 0.0000104658058052988)
};\addlegendentry{(A): emp., $m = 10$};
 
\addplot[red, line width=0.7pt]coordinates{
(2, 1.00000000000000)
(4, 1.00000000000000)
(6, 1.00000000000000)
(8, 0.958712838843798)
(10, 0.374955030679825)
(12, 0.0584574595968143)
(14, 0.00657642332608339)
(16, 0.000647346704152758)
(18, 0.0000757224846858924)
(20, 7.13145866478082e-6)
};\addlegendentry{(A): th., $m = 10$};

\addplot[only marks, mark=+,  black, mark size=2.5pt,line width=0.7pt]coordinates{
(2, 1.00000000000000)
(4, 1.00000000000000)
(6, 0.840336134453782)
(8, 0.260416666666667)
(10, 0.0347463516330785)
(12, 0.00617703378837482)
(14, 0.000687866718944537)
(16, 0.0000934658928936947)
(18, 0.0000128284170990485)
(20, 1.47618506070014e-6)
};\addlegendentry{(A): emp., $m = 20$};
 
\addplot[black, line width=0.7pt]coordinates{
(2, 1.00000000000000)
(4, 0.999999966145167)
(6, 0.871884238333734)
(8, 0.246032782325948)
(10, 0.0357900777670877)
(12, 0.00507354929663443)
(14, 0.000660509770270156)
(16, 0.0000844310355451482)
(18, 0.0000112323348099458)
(20, 1.31895461649768e-6)
};\addlegendentry{(A): th., $m = 20$};

\addplot[only marks, mark=+,cyan, mark size=2.5pt,line width=0.7pt]coordinates{
(2, 1.00000000000000)
(4, 1.00000000000000)
(6, 0.757575757575758)
(8, 0.211864406779661)
(10, 0.0570450656018254)
(12, 0.00976944118796405)
(14, 0.00152144476394784)
(16, 0.000226399032823332)
(18, 0.0000340374895717641)
(20, 4.96241344000206e-6)
};\addlegendentry{(A): emp., $m = 100$};
 
\addplot[cyan, line width=0.7pt]coordinates{
(2, 1.00000000000000)
(4, 0.997629111225400)
(6, 0.692728118579051)
(8, 0.215446835142498)
(10, 0.0463750758168842)
(12, 0.00874542032658920)
(14, 0.00152700843404913)
(16, 0.000248740223380801)
(18, 0.0000376788759471895)
(20, 5.27580799335954e-6)
};\addlegendentry{(A): th., $m = 100$};

\addplot[only marks, mark=+,mygreen, mark size=2.5pt,line width=0.7pt]coordinates{
(2, 1.00000000000000)
(4, 1.00000000000000)
(6, 1.00000000000000)
(8, 1.00000000000000)
(10, 1.00000000000000)
(12, 1.00000000000000)
(14, 1.00000000000000)
(16, 0.961538461538462)
(18, 0.833333333333333)
(20, 0.571428571428571)
(22, 0.311526479750779)
(24, 0.186915887850467)
(26, 0.0991080277502478)
(28, 0.0577034045008656)
(30, 0.0312012480499220)
(32, 0.0146412884333821)
 (34, 0.00923616883716634)
 (36, 0.00493023714440665)
 (38, 0.00287902343525076)
(40, 0.00135582189923532)
(42, 0.000730695037119308)
(44, 0.000462821545268575)
(46, 0.000221955875172016)
(48, 0.000125401755875386)
};\addlegendentry{(B): emp., $m = 200$};
 
\addplot[mygreen, mark size=2pt,line width=0.7pt]coordinates{
(2, 1.00000000000000)
(4, 1.00000000000000)
(6, 1.00000000000000)
(8, 1.00000000000000)
(10, 1.00000000000000)
(12, 0.999999997279301)
(14, 0.999746835275917)
(16, 0.974573103684654)
(18, 0.816062737276230)
(20, 0.567053420393270)
(22, 0.342462010595671)
(24, 0.193527200911166)
(26, 0.105385701250370)
(28, 0.0568531687268642)
(30, 0.0309301784350855)
(32, 0.0166320860184482)
 (34, 0.00917495653263975)
 (36, 0.00504546806291021)
 (38, 0.00266293844985346)
 (40, 0.00145496338207622)
 (42, 0.000793569105105269)
 (44, 0.000432125160333640)
(46, 0.000234937683582129)
(48, 0.000127534314880529)
};\addlegendentry{(B): th., $m = 200$};

\addplot[only marks, mark=+,blue, mark size=2.5pt,line width=0.7pt]coordinates{
 (2, 1.00000000000000)
 (4, 1.00000000000000)
 (6, 1.00000000000000)
 (8, 1.00000000000000)
 (10, 1.00000000000000)
 (12, 1.00000000000000)
 (14, 1.00000000000000)
 (16, 0.980392156862745)
 (18, 0.862068965517241)
 (20, 0.675675675675676)
 (22, 0.490196078431373)
 (24, 0.271739130434783)
 (26, 0.204081632653061)
 (28, 0.0966183574879227)
 (30, 0.0545851528384279)
 (32, 0.0302938503483793)
 (34, 0.0154511742892460)
 (36, 0.0106871860639094)
 (38, 0.00576501787155540)
(40, 0.00284997720018240)
(42, 0.00148725423123829)
(44, 0.000966090232827746)
(46, 0.000403160780519271)
(48, 0.000266886581209583)
};\addlegendentry{(B): emp., $m = 400$};
 
\addplot[blue, mark size=2pt,line width=0.7pt]coordinates{
 (2, 1.00000000000000)
 (4, 1.00000000000000)
 (6, 1.00000000000000)
 (8, 1.00000000000000)
 (10, 0.999999999998425)
 (12, 0.999998838683034)
 (14, 0.999284327015783)
 (16, 0.979968906824613)
 (18, 0.881642587064298)
 (20, 0.686336599466160)
 (22, 0.473201241123758)
 (24, 0.298064213029340)
 (26, 0.177312555050825)
 (28, 0.101919155926849)
 (30, 0.0574077095985928)
 (32, 0.0319489503359565)
 (34, 0.0176501984063684)
 (36, 0.00970488186421758)
 (38, 0.00531878565851922)
 (40, 0.00290780984570925)
 (42, 0.00158650845828596)
 (44, 0.000864063588513086)
(46, 0.000469820171449091)
(48, 0.000255052364759586)
};
\addlegendentry{(B): th., $m = 400$};

\addplot[only marks, mark=+,orange, mark size=2.5pt,line width=0.7pt]coordinates{
 (2, 1.00000000000000)
 (4, 1.00000000000000)
 (6, 1.00000000000000)
 (8, 1.00000000000000)
(10, 1.00000000000000)
 (12, 1.00000000000000)
 (14, 1.00000000000000)
 (16, 1.00000000000000)
 (18, 0.980392156862745)
 (20, 0.943396226415094)
 (22, 0.763358778625954)
 (24, 0.571428571428571)
 (26, 0.420168067226891)
 (28, 0.264550264550265)
 (30, 0.136425648021828)
(32, 0.0674308833445718)
 (34, 0.0483091787439614)
 (36, 0.0232558139534884)
 (38, 0.0135685210312076)
(40, 0.00839066957543212)
(42, 0.00356862465205910)
(44, 0.00217159981758562)
(46, 0.00111270598969634)
(48, 0.000667748419105618)
};\addlegendentry{(B): emp., $m = 1000$};
 
\addplot[orange, mark size=2pt,line width=0.7pt]coordinates{
(2, 1.00000000000000)
(4, 1.00000000000000)
(6, 1.00000000000000)
(8, 1.00000000000000)
(10, 1.00000000000000)
(12, 0.999999999999964)
(14, 0.999999962647955)
(16, 0.999922420780881)
(18, 0.994693547241745)
(20, 0.944898982808872)
 (22, 0.798575702446249)
 (24, 0.587196413542887)
 (26, 0.386114784825158)
 (28, 0.235656489515202)
 (30, 0.137399494723810)
 (32, 0.0779687290782790)
 (34, 0.0435431003068234)
 (36, 0.0240858942767295)
 (38, 0.0132439684214838)
 (40, 0.00725367850393723)
(42, 0.00396155300181466)
(44, 0.00215875928686856)
(46, 0.00117413659041928)
(48, 0.000637508945130060)
};\addlegendentry{(B): th., $ m = 1000$};

\end{axis}
\end{tikzpicture}%
    }
    \caption{Simulated values of $\epsilon$ and comparison with its theoretical values from \eqref{eq:epsilon}. The following two settings are considered: (A) $(n,\gamma) = (100,\frac{1}{2})$, (B) $(n,\gamma) = (1000,\frac{3}{4})$.}
    \label{fig:sim_comparison}
\end{figure}

First of all, we consider $\beta = 1$, \ie, we assume that malicious full nodes hide the whole content of a block.
In fact, from the analysis in Section \ref{buterin}, it is clear that the most convenient strategy for malicious nodes is to reveal symbols only when queried by light nodes.
Starting from this observation, the analysis of the ASBK protocol can be significantly simplified, by employing some basic approximations.
\begin{Prop}\label{prop:probability}
Let us consider the ASBK protocol with $\beta=1$, employing a 2D-RS code with length $n$ and able to fill up to $(1-\gamma) n$ erasures. 
Let $m$ be the number of light nodes, each querying a malicious node by asking for $s$ random and distinct symbols.
Then, the adversarial success probability can be well approximated as
\begin{equation}
\epsilon = \begin{cases}1&\text{if $x^*<\gamma n$,}\\
1-\left(1-\frac{\binom {\gamma n -1 }{s}}{\binom{x^*}{s}}\right)^m&\text{otherwise,}\end{cases}
\label{eq:epsilon}
\end{equation}
where $x^* = n\left(1-\left(1-\frac{s}{n}\right)^m\right).$
\end{Prop}
\begin{IEEEproof}
We assume that, in every execution of the protocol, the number of distinct symbols  requested by the light nodes altogether is equal to its expected value that, recalling \eqref{eq:mean_t} with the appropriate notation, can be estimated as
\begin{equation}
x^* = \langle x \rangle = n\left(1-\left(1-\frac{s}{n}\right)^m\right).
\label{eq:xstardef}
\end{equation}
We remind that at least $\gamma n$ symbols are necessary to decode the employed 2D-RS code.
Hence, when $x^*<\gamma n$, we can set $\mathrm{Pr}[\textsf{NoDec}] = 1$ and, recalling \eqref{eq:eps}, we get $\epsilon = 1$. 
When instead $x^*\geq\gamma n$, decoding is always successful, and, consequently, we set $\mathrm{Pr}[\textsf{NoDec}] = 0$.
In such a case, the best possible strategy for the adversary is to ignore some queries, keeping $d$ symbols hidden.
In order to prevent decoding, it must be $d > x^*-\gamma n$; hence, in order to maximize the success probability, the adversary sets $d = x^*-\gamma n +1$.

We know that the ensemble of light nodes asks for a total of $x^*$ symbols; let $J\subseteq \{1,\ldots,n\}$ be the set of positions pointing at such $x^*$ symbols. We can consider that each light node randomly selects $s$ distinct indexes of $J$.
Let $D\subseteq J$ be the set of indexes pointing at the $d$ positions of the symbols that the malicious node keeps hidden: a light node will not put the block in pending state if and only if it only requests positions coming from $J\setminus D$.
We now proceed by computing the probability that there is at least a light node that only requests symbols in positions indexed by $J\setminus D$, whose size is $x^*-d$.
The probability that a single light node asked for a symbol in $D$ is $p = 1-\binom{x^*-d}{s}/\binom{x^*}{s}$. Note that $p$ corresponds to the probability that a single light node is not misled, \ie, does not end up accepting the block.
Since there are $m$ light nodes operating independently, we have that the adversarial success probability, \ie, the probability that there is at least one light node that did not select positions from $D$, is obtained as the complementary of the probability that every light node asked at least a symbol coming from $D$.
Then
\begin{equation*}
\epsilon = 1-p^m =  1-\left(1-\frac{\binom{x^*-d}{s}}{\binom{x^*}{s}}\right)^m.    
\end{equation*}
Considering $d = x^*-\gamma n +1$, we prove the thesis.
\end{IEEEproof}
In order to confirm the validity of our analysis, we have run numerical simulations of the ASBK sampling process. 
We have measured the adversarial success probability and compared it with the ones obtained through Proposition \ref{prop:probability}; results are shown in Fig. \ref{fig:sim_comparison}. We observe a good matching between the analytical values and the simulated ones.

\subsection{Model Introspection and Asymptotic Analysis}



Let us write $x^* = n(1-\sigma)$, with $\sigma \in [\![ 0 ; 1 ]\!]$, and  obtain from \eqref{eq:xstardef} that $ \ln \sigma = m\ln \left(1-\frac{s}{n}\right)   
$.
Since it is desirable to have $s\ll n$, we have $\ln\left(1-\frac{s}{n}\right)\approx - \frac{s}{n}$, from which $
ms \approx n\ln(1/\sigma)$.  We want full nodes to be able to decode; hence, taking into account \eqref{eq:gamma}, it must be $\sigma \leq 1 - \gamma \approx 1 - R'(2-R')$.
We can then derive a lower bound on the product $ms$ as
\begin{equation} 
\label{eq:s_lower_bound}
ms > n\ln\left(\frac{1}{1 - \gamma}\right) \approx n\ln\left(\frac{1}{1 - R'(2-R')}\right).
\end{equation}
The above condition gives a rough idea of the setting one should choose in ASBK.
Indeed, we have that when \eqref{eq:s_lower_bound} is not satisfied, then, with high probability, full nodes will not be able to decode.
Additionally, \eqref{eq:s_lower_bound} explicitly expresses the relation between the component code rate and the total number of samples which are asked for full nodes (that is, the value of $ms$).
If one wants to keep the value of $s$ as low as possible, assuming that the number $m$ of light nodes is fixed, then the only possibility consists in reducing the code rate $R'$.
Notice that, by doing this, one also reduces the value of $\gamma$, which is the reason why a smaller number of queries is required: the obtained 2D-RS code can fill a larger number of erasures.

As we know, the number of nodes participating in modern blockchain networks is rapidly increasing.
This implies that the number of light nodes is increasing as well; so, it is worthwhile  analyzing the ASBK protocol also in this regime, \ie, when $m$ becomes extremely large.
In particular, as $m$ grows, from \eqref{eq:xstardef} we see that $x^*$ rapidly tends to $n$.
In the limit of $x^* = n$, from Proposition \ref{prop:probability} we obtain
$$\epsilon = 1-\left(1-\frac{\binom{\gamma n -1}{s}}{\binom{n}{s}}\right)^m.$$
We observe that, whenever $s\ll (1-\gamma)n$, we can set
$\binom{\gamma n - 1}{s}/\binom{n}{s}\approx \gamma^s$. 
Exploiting such an approximation, we find  
\begin{equation}
\label{eq:medium_s_approx}
s \approx \frac{\ln\left(1-(1-\epsilon)^{\frac{1}{m}}\right)}{\ln\gamma}.    
\end{equation}
Since $\epsilon$ is relatively small and $m$ is relatively large, we can make use of the following further approximations
$$(1-\epsilon)^{\frac{1}{m}}\approx 1+\frac{\ln(1-\epsilon)}{m},\hspace{4mm}\ln(1-\epsilon)\approx -\epsilon,$$
which permit us to rewrite \eqref{eq:medium_s_approx} as 
\begin{equation}
\label{eq:s_approx}
s \approx \frac{\ln (m) +\ln\left(\frac{1}{\epsilon}\right)}{\ln\left(\frac{1}{R'(2-R')}\right)}.
\end{equation}
We first notice that increasing the block size has basically no impact on the value of $s$ which is required to reach a desired adversarial success probability.
However, the value of $s$ grows with the network size (which is indirectly measured by $m$), but the growth rate is only logarithmic.

\section{Results}
\label{results}
In this section we validate our analysis, finding optimal settings for the ASBK protocol\footnote{The software programs used to obtain the results in this paper are available at \url{https://github.com/secomms/blockchainRS}}. 
In order to consider a case of practical interest, we choose block sizes which are similar to those used in common blockchains (such as Ethereum), and find the optimal protocol setting for different network sizes. 

We first define the link between code parameters and block size, assumed to be equal to $\ell_b$ bits. 
Considering a code defined over $\mathbb F_q$, we need to use component RS codes with dimension 
$$k' = \left\lceil\sqrt{\frac{\ell_b}{\log_2{q}}}\hspace{2mm}\right\rceil,$$ 
which guarantee that the list of transactions fits into a square block with side $k'$ (if needed, padding is employed to fill all the matrix entries). We now proceed by determining the header size.
We exclude from our analysis all the header elements which do not depend on the code design, such as the hash of the header of the previous block, since they provide a constant contribution which is the same in all the considered cases.
Consequently, we assume that the header only contains the Merkle roots of the encoded block, which are $2n'$.
By denoting with $\ell_\hash = 256$ the binary length of the digests, we have that the header size is
$2n'\ell_\hash$.
We now consider that the reply to each
light node query is composed by an element of $\mathbb F_q$ and its Merkle proof.
A Merkle proof has size $\lceil\log_2(n')\rceil\ell_\hash$, while each symbol of $\mathbb F_q$ is represented by $\lceil\log_2(q)\rceil$ bits. 
Hence, a light node downloads a total amount of data, in bits, given by
$$\ell_D = 2 \ell_\hash k'/R' + s\big(\lceil\log_2(q)\rceil + \ell_\hash\left\lceil\log_2\left(k'/R'\right)\right\rceil\big).$$
In order to provide parameters with practical interest, we consider a block size of 75 kB\footnote{This is the average block size of the Ethereum network in August 2021. Data are extracted from Etherscan (\url{https://etherscan.io/chart/blocksize}).}, yielding $\ell_b = 600,000$. 

\begin{figure}
    \centering
    \resizebox{!}{5.1cm}{%
    \input{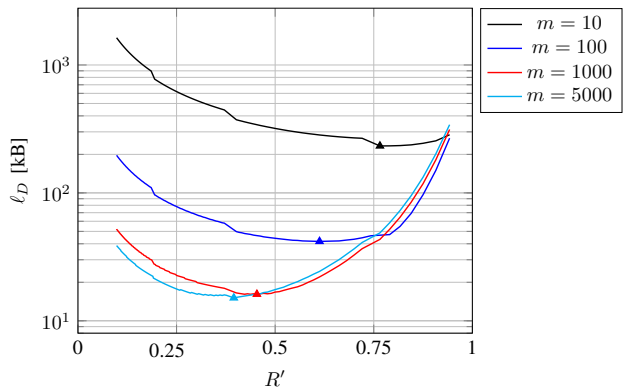}%
    }
    \caption{Amount of downloaded data as a function of the component code rate $R'$, when $q = 2^{256}$. 
    }
    \label{fig:data_size_vs_rate}
\end{figure}

Firstly, we consider $q = 2^{256}$ and several values of $R'$, and for each configuration we find the minimum value of $s$ for which the adversarial success probability is below the target $0.01$ (as in \cite{asbk21}). 
The corresponding values of $\ell_D$ are shown in Fig. \ref{fig:data_size_vs_rate}, where we consider different values of $m$ and, for each curve, we highlight the value of $m$ yielding the minimum $\ell_D$.
We observe how the amount of downloaded data decreases for low code rates, reaches its minimum and then increases for higher code rates.
Moreover, for low rates, the number of samples (or downloaded data) decreases as $m$ increases, while the opposite tends to occur for high rate values.
As expected, the optimum working point depends on $m$: in a real network, the value of $R'$ should be adjusted as the network size changes.

Note that, as another degree of freedom, one may change $q$.
In Table \ref{table:min_D_1000} we have reported the optimal rate values, found for several values of $q$ when $m = 1000$.
To quantify the gain with respect to the case of $R' = 0.5$, the last two columns report the values of $\ell_D$ and $s$ when $R' = 0.5$ (which we have denoted, respectively, as $\tilde{\ell}_D$ and $\tilde{s}$).
As we see from the table, for all the considered cases, the optimum rate never corresponds to $0.5$.
We further notice that the finite field size $q$ plays a crucial role in determining the amount of data each light node downloads.
Indeed, it appears that using a rather large finite field makes the value of $\ell_D$ decrease significantly. For example, in our scenario, $\ell_D$ achieves its minimum for $q=2^{2048}$. 
At a first glance, such a large $q$ may appear unpractical.
However, notice that sums and multiplications in $\mathbb F_q$ cost $O\big(\log_2(q)\big)$ and $O\big(\log_2^2(q)\big)$, respectively. For instance, $\mathbb F_{2^{2048}}$ is $2^{1792}$ times larger than $\mathbb F_{2^{256}}$, but sums and multiplications in $\mathbb F_{2^{2048}}$ are expected to cost only $8$ and $64$ times as much than in $\mathbb F_{2^{256}}$, respectively.
Furthermore, some specific choices for the finite field construction may lead to computational advantages (see \cite{maximov2017fast}, for instance).
In any case, in a practical situation, the choice on $q$ should be based (also) on the computational resources actually available to full nodes.

\begin{table}[t!]
\centering
\caption{Optimal rates and corresponding values of $\ell_D$ and $s$, for $m = 1000$ and $\epsilon = 10^{-2}$.}
\begin{tabular}{ | c c | c c c c| c c| } 
    \hline    &  &  &  & &  &  & \\[-8pt]
    $q$ & $k'$ & $R'$ & $n'$ & $\ell_D$ & $s$ & $\tilde{\ell}_D$ & $\tilde{s}$ \\
    \hline
    $2^{16}$ & 194 & 0.647  & 300 & 84.740 & 226 & 92.112 & 232\\
    $2^{32}$ & 137 & 0.591 & 232 & 48.128 & 136 & 54.912 & 128\\
    $2^{64}$ & 97 & 0.545 & 194 & 31.984 & 78 & 32.480 & 76\\
    $2^{128}$ & 69 & 0.548 & 126 & 21.504 & 56 & 22.704 & 51\\
    $2^{256}$ & 49 & 0.430 & 114 & 16.256 & 35 & 16.768 & 41\\
    $2^{512}$ & 35 & 0.402 & 87 & 13.344 & 27 & 15.424 & 38\\
    $2^{1024}$ & 25 & 0.321 & 78 & 11.680 & 19 & 15.040 & 37\\
    $2^{2048}$ & 18 & 0.295 & 61 & 11.072 & 16 & 17.984 & 35\\
    $2^{4096}$ & 13 & 0.224 & 58 & 12.160 & 12 & 23.840 & 33\\
    $2^{8192}$ & 9 & 0.155 & 58 & 14.656 & 10 & 36.672 & 30\\
     \hline
\end{tabular}
\label{table:min_D_1000}
\end{table}

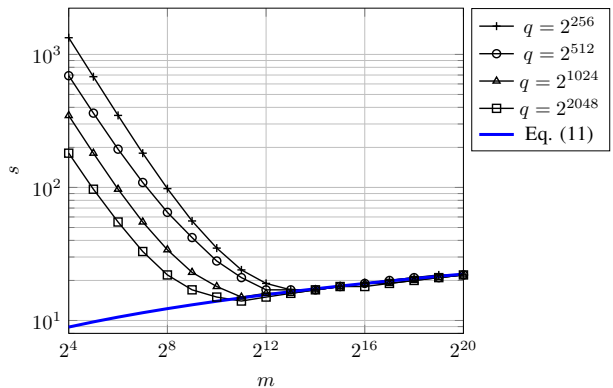
\begin{figure}
    \centering
    \resizebox{!}{5.1cm}{%
    \begin{tikzpicture}
  \begin{axis}[
xtick={2^4,2^8,2^12,2^16,2^20},        
xticklabels={$2^4$,$2^8$,$2^{12}$,$2^{16}$,$2^{20}$},        
xmin = 16,
xmax= 2^20,
grid = both,
ymin=8,
xmode = log,
ymode = log,
legend style={at={(1.02,1)},anchor=north west},
mark size=2pt,
xlabel={$\textcolor{white}{R'}m\textcolor{white}{R'}$},
ylabel={$s$},
xticklabel style={
        /pgf/number format/fixed,
        /pgf/number format/precision=5
}]

\addplot[black, mark = +, line width=0.7pt]coordinates{
(16, 1335)
(32, 680)
(64, 348)
(128, 181)
(256, 98)
(512, 56)
(1024, 35)
(2048, 24)
(4096, 19)
(8192, 17)
(16384, 17)
(32768, 18)
(65536, 19)
(131072, 20)
(262144, 21)
(524288, 22)
(1048576, 22)

};\addlegendentry{$q=2^{256}$};

\addplot[black, mark = o, line width=0.7pt]coordinates{
(16,692)
(32,362)
(64,194)
(128,109)
(256,65)
(512,42)
(1024,28)
(2048,21)
(4096,17)
(8192,17)
(16384,17)
(32768,18)
(65536,19)
(131072,20)
(262144,21)
(524288,21)
(1048576,22)
};\addlegendentry{$q=2^{512}$};

\addplot[black, mark = triangle, line width=0.7pt]coordinates{
(16, 347)
(32, 181)
(64, 97)
(128, 55)
(256, 34)
(512, 23)
(1024, 18)
(2048, 15)
(4096, 16)
(8192, 16)
(16384, 17)
(32768, 18)
(65536, 19)
(131072, 19)
(262144, 20)
(524288, 21)
(1048576, 22)
};\addlegendentry{$q=2^{1024}$};

\addplot[black, mark = square, line width=0.7pt]coordinates{
(16, 181)
(32, 97)
(64, 55)
(128, 33)
(256, 22)
(512, 17)
(1024, 15)
(2048, 14)
(4096, 15)
(8192, 16)
(16384, 17)
(32768, 18)
(65536, 18)
(131072, 19)
(262144, 20)
(524288, 21)
(1048576, 22)
};\addlegendentry{$q=2^{2048}$};

\addplot[blue,  line width=1.5pt]coordinates{
(16, 8.92457981559608)
(32, 9.76305223165237)
(64, 10.6015246477087)
(128, 11.4399970637649)
(256, 12.2784694798212)
(512, 13.1169418958775)
(1024, 13.9554143119338)
(2048, 14.7938867279901)
(4096, 15.6323591440464)
(8192, 16.4708315601027)
(16384, 17.3093039761590)
(32768, 18.1477763922152)
(65536, 18.9862488082715)
(131072, 19.8247212243278)
(262144, 20.6631936403841)
(524288, 21.5016660564404)
(1048576, 22.3401384724967)

};\addlegendentry{Eq. \eqref{eq:s_approx}};


\end{axis}
\end{tikzpicture}%
    }
    \caption{Minimum number of samples to achieve $\epsilon \leq 10^{-2}$, as a function of $m$; the code rate is $R'=0.25$.}
    \label{fig:samples_vs_m}
\end{figure}

Finally, we study how the number of asked samples varies as $m$ grows, for several values of $q$.
The obtained results are reported in Fig. \ref{fig:samples_vs_m}, where we also show the values resulting from \eqref{eq:s_approx}.
As expected, as $m$ grows, the values of $s$ tend to become closer and closer to those given by \eqref{eq:s_approx}.
Moreover, as correctly predicted by \eqref{eq:s_approx}, $s$ does not depend on $q$ and $n$ and, therefore, the block size has no impact. In other words, $s$ is only a function of $R'$ and $m$.
This implies that, for very heavily participated networks, one can increase the amount of data in each block without any evident effect on the computational burden of each light node.

\section{Conclusions}
\label{conclusions}
We have presented a novel mathematical model for the ASBK protocol, a recently proposed blockchain protocol which counters data availability attacks through 2D-RS codes.  
Differently from existing analyses, our study does not fix any code parameter and embeds easy-to-implement formulas. 
This allows for a deeper understanding of the protocol features, and ultimately provides a simple method to devise optimal settings.
Namely, our approach allows to settle the best code parameters (e.g., rate and finite field size) to minimize the amount of data each light node downloads, given a desired adversarial success probability.
Our results show that the ASBK protocol benefits from the use of component codes with rate different from the value $1/2$, fixed in the original proposal.


\bibliographystyle{IEEEtran}
\bibliography{Archive}


\end{document}